\documentclass[twocolumn,preprintnumbers,amsmath,amssymb]{revtex4}

\usepackage{graphicx}  
\begin{document}
\title{
Injection terahertz laser using the resonant inter-layer radiative transitions
in   double-graphene-layer structure
}
\author{V.~Ryzhii$^{1,2}$,
A.~A.~Dubinov$^{3}$, V.~Ya.~Aleshkin$^{3}$,   M.~Ryzhii$^{2,4}$, 
and T.~Otsuji$^{1,2}$
}
\affiliation{
$^1$Research Institute for Electrical Communication, Tohoku University, Sendai 980-8577, Japan\\
$^2$ Japan Science and Technology Agency, CREST, Tokyo 107-0075, Japan\\
$^3$Institute for Physics  of Microstructures of Russian Academy of Sciences, and Lobachevsky State University of Nizhny Novgorod, Nizhny Novgorod 603950, Russia\\
$^4$Department of Computer Science and Engineering, University of Aizu, Aizu-Wakamatsu  965-8580, Japan\\
}
\begin{abstract}
We propose and substantiate the concept of  terahertz (THz) laser  enabled by the resonant  electron radiative  transitions between graphene layers (GLs) in double-GL structures.
We estimate the THz gain for TM-mode exhibiting very low Drude absorption in GLs  and show that the gain  can exceed the losses in metal-metal waveguides at the low end of the THz range. The spectrum of the emitted photons can be tuned by the applied voltage.
A weak temperature dependence of the THz gain promotes an effective operation at room temperature. 
\end{abstract}

\maketitle

The gapless energy spectrum of graphene layers (GLs)~\cite{1} enables the creation of different
terahertz (THz) devices utilizing the interband transition. In particular,  the interband population 
inversion and the pertinent negativity of the dynamic conductivity in GLs~\cite{2,3} due to the optical or 
injection pumping can be used in
GL-based THz lasers~\cite{4,5,6,7,8,9}. 
First experimental results on the THz emission from optically excited GLs~\cite{10}
(see also review paper~\cite{11} and references therein) instill confidence in the realization of such lasers.
One of the obstacles, limiting the achievement of the negative dynamic conductivity in the range of a few THz,   
is the reabsorption of the photons with the in-plane polarization
emitted at the interband transitions 
due to the intraband transitions (the Drude absorption). 
Similar situation takes place in the quantum cascade lasers (QCLs) based on multiple quantum well (MQW)  
structures~\cite{12}.
However, in the case of the photon polarization perpendicular to the QW plane the intraband
(intrasubband) absorption can be much weaker than that following from the semi-classical Drude formula~\cite{13}.

In this paper, we propose a device structure based on a double-GL structure shown in Fig.~1
(upper panel)  with the injection
of electrons to one n-doped GL and  to another p-doped GL, which can be used for lasing of THz photons 
with the electric field perpendicular to the GL plane
due to the tunneling inter-GL radiative processes. The structure band diagram under the applied bias voltage 
and tunneling transitions assisted with the emission of photons 
with the energy $\hbar\omega \sim \Delta$, where $\Delta$ is the energy distance (gap) between 
the Dirac points in GLs,
 are demonstrated in Fig.~1 (lower panel). These transitions take place from the conduction band
of the GL with 2DEG to the empty conduction band of GL with 2DHG. The transition from
the filled valence band the former GL   to the empty portion of the valence band of the latter 
GL also  contribute to the emission of photons with $\hbar\omega \sim \Delta$.
The structure comprises two GLs with the side contact at one of the GL edges.
This double-GL structure plays the role of the laser active region.
The opposite edge of GL is isolated from another contact. A narrow  tunneling-transparent
barrier separates GLs (its thickness $d$ is about few nanometers). The applied bias voltage $V $provides  
 the formation
of the two-dimensional electron and hole gases (2DEG and 2DHG) in the upper and lower GLs, respectively 
owing to the injection from the side contacts, so that the inter-GL population inversion occurs.
The GL structures in question were fabricated recently~\cite{14,15,16,17}. The effective control of the GL 
population was,
in particular, used for modulation of optical radiation~\cite{14,15}. These structures can exhibit
marked inter-GL tunneling  current, which in the case of the Dirac point alignment,
exhibit the negative differential inter-GL conductivity~\cite{18} (see also Refs.~\cite{19,20,21}, where 
the latter effect was theoretically considered). 
The inter-GL barrier layers can be made of hBN, WS$_2$, or similar materials. 
The double-GL structures can also be used in different devices (THz detectors and photomixers) utilizing 
the plasmonic effects~\cite{22,23,24}.

The THz laser metal-metal (MM) waveguide system consist of two parallel metal strips
(see Fig.~1, upper panel) as in some QCLs~\cite{12}. The net spacing between the strips $(2W + d)$
is about $10~\mu$m. The spatial distributions of the amplitudes of THz electric-field components $|E_z|$ 
and $|E_x|$ in the TM-mode propagating in the $y$-direction
 and the real part of refractive index in the MM waveguide are shown in Fig.~2.
It is assumed that the device structure is supplied with the proper mirrors reflecting the radiation 
propagating in the $y$-direction.

To enhance the laser output power, a more complex active region can used which includes several double-GL 
structures (with thin inter-GL layers and relatively thick layers separating the double-GL structures) 
or a multiple-GL structure with the tunneling-transparent inter-GL  barrier layers.
A wave guide system using the specific plasmon-polaritons~\cite{25} (see also~Ref.\cite{7}), associated 
the high conductivity of GLs and propagating along the structure (in the $y$
-direction) can also be implemented.

Apart from weaker intraband reabsorption of the generated photons, the double-GL structures
(and more complex ones) might exhibit the advantages in comparison with
the p-i-n lasers with simultaneous injection of both electrons and holes into the same GLs~\cite{8,26,27}
because of a  more effective injection due to the absence of the electron-hole "friction"~\cite{28}
and the weakening of nonradiative recombination mechanisms, in particular, that associated with the emission 
of optical phonons.

\begin{figure}[t]\label{Fig.1}
\begin{center}
\includegraphics[width=5.0cm]{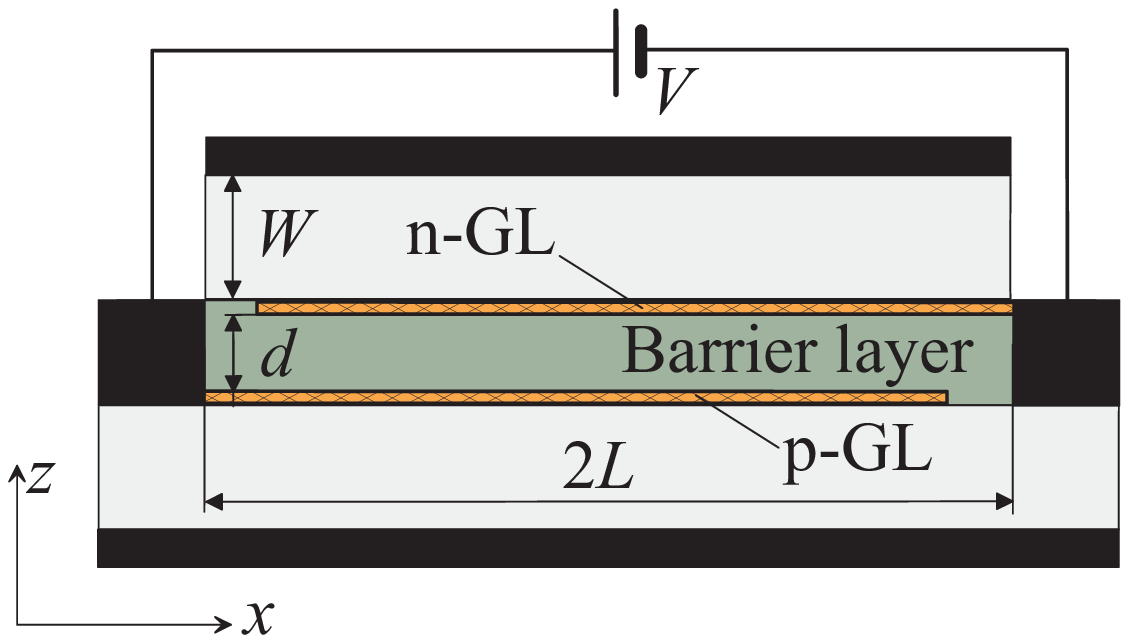}\\
\vspace*{0.4cm}
\includegraphics[width=5.0cm]{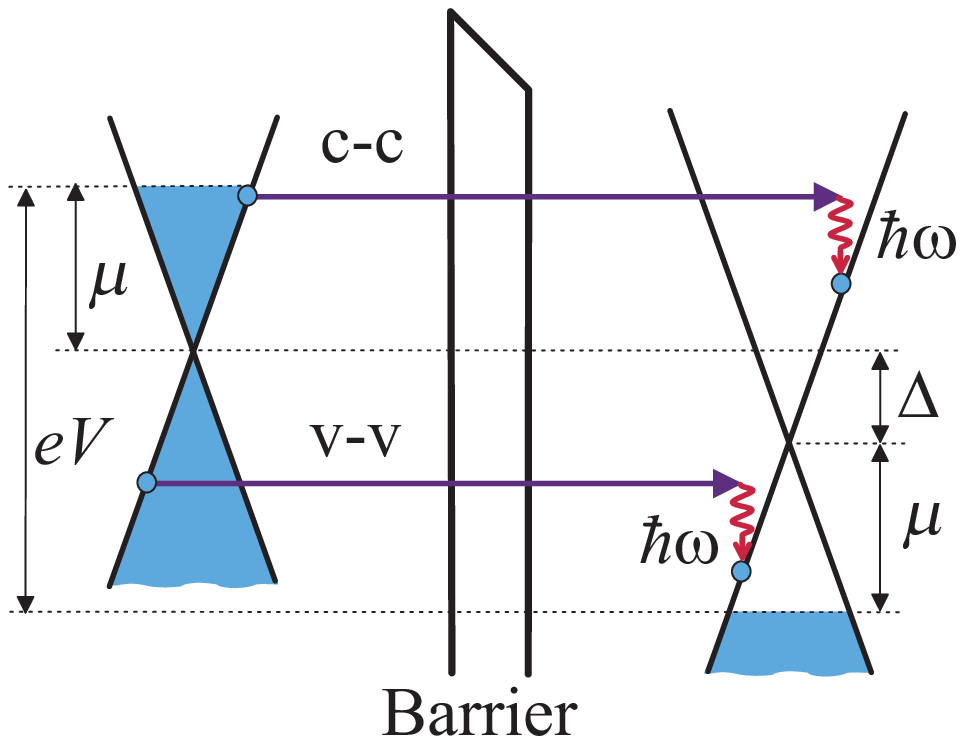}\\
\caption{
Schematic view of the DG-laser  with double-GL structure  and MM waveguide(upper panel)
and the device band diagram (lower panel). Dark regions correspond to the highly conducting side contacts 
and metal wave-guide strips (along the $y$-direction. 
Arrows in the lower panel indicate the resonant-tunneling (with the conservation of the electron momentum) 
inter-GL transitions between the conduction band  ($c-c$) and the valence band ($v-v$) states and 
assisted by the emission of photons (at weak depolarization shift).}
\end{center} 
\end{figure}

\begin{figure}[t]\label{Fig.2}
\begin{center}
\includegraphics[width=7.0cm]{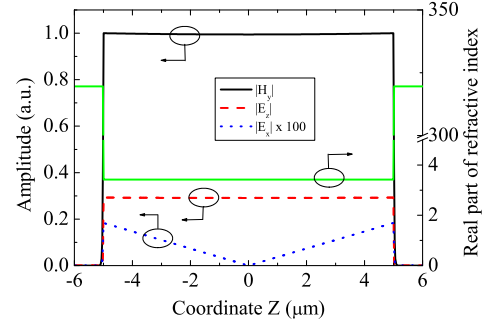}\\
\caption{Spatial distributions of the amplitudes of THz electric- and magnetic-field components
in TM mode and the real part of the refraction index in the MM waveguide under consideration.
}
\end{center} 
\end{figure}

\begin{figure}[t]\label{Fig.3}
\begin{center}
\includegraphics[width=5.5cm]{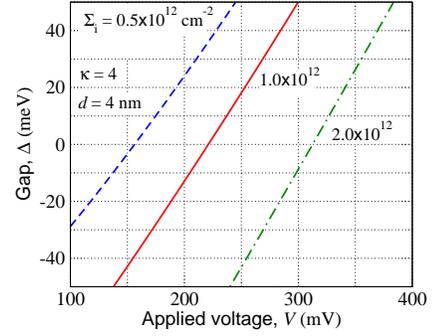}
\caption{Energy gap between Dirac points  $\Delta$ versus applied voltage $V$ at different dopant 
densities $\Sigma_i$.
}
\end{center} 
\end{figure}

The electron and hole density in the pertinent GLs is given by

\begin{equation}\label{eq1}
\Sigma = \Sigma_i + \frac{\kappa\Delta}{4\pi\,e^2d},
\end{equation}
where $\Sigma_i$ is the density of donors in the upper GL (n-GL) and acceptors in the lower GL (p-GL),  
 $\kappa$ is the dielectric constant, and $e = |e|$ is the electron charge.  
If the length of the electron and hole diffusion  exceeds the length of GLs, the local potential 
difference between GLs is close to $V$~\cite{8,27}. In this situation, the quantity $\Delta$ 
(which determines the electric field $E = \Delta/ed$ in the inter-GL barrier)  obeys the following equation (see Fig.~1):

\begin{equation}\label{eq2}
\Delta + 2\mu = eV,
\end{equation}
where $\mu$ is the value of the Fermi energy of 2DEG and 2DHG in the pertinent GL.
In the case of the 2DEG and 2DHG    strong degeneration ($\mu \gg T$, where $T$ is the temperature 
in the energy units), 

\begin{equation}\label{eq3}
 \mu \simeq \hbar\,v_W\sqrt{\pi\Sigma},
\end{equation}
respectively, where $v_W \simeq 10^8$~cm/s is the characteristic velocity of electrons and holes in GLs.
Considering  Eqs.~(1) - (3), in the above limiting cases, we arrive at 
\begin{equation}\label{eq4}
 \Delta/e = V + V_0 -\sqrt{2VV_0 + V_0^2 + V_t^2}, 
\end{equation}
where $V_0 = \hbar^2v_W^2\kappa/2e^3d$ and $V_t = 2\hbar\,v_W\sqrt{\pi\Sigma_i}/e$
At $V \leq V_t$, $\Delta \leq 0 $, while at $V > V_t$,  $\Delta > 0 $.
The band diagram at the bias voltage  $V > V_t$ is  shown in Fig.~1 (lower panel).
In this case, the tunneling inter-GL transitions are possible only due to the scattering processes
accompanying the tunneling~\cite{19,20,21}. However, the inter-GL transitions assisted by the emission of 
photons with the polarization corresponding to the photon electric field
perpendicular to GLs (along the axis $z$) conserve the electron momentum and, hence,
do not require any scattering (resonant-tunneling photon-assisted transitions).
Assuming $\kappa = 4$, $d = 4$~nm, and $\Sigma_i = 10^{12}$~cm$^{-2}$, one obtains
$V_0 \simeq 136$~mV and $V_t \simeq 221$~mV. The quantities $\Delta = 5 - 10$~meV
correspond to $V \simeq 229 - 237$~mV and $\mu \simeq 112.0 - 113.5$~meV.

The real part of the transverse  ac conductivity, Re~$\sigma_{zz}(\omega)$, of the double-GL structure 
under consideration 
can be estimated using the following formula:
  
\begin{equation}\label{eq5}
{\rm Re}~\sigma_{zz}(\omega) = - \frac{2e^2}{\hbar}\frac{|z_{u,l}|^2\Sigma_i
(1 + \Delta/\Delta_i)\gamma\hbar\omega}{[\hbar^2(\omega - \omega_{max})^2 + \gamma^2]}.
\end{equation}
Here 
\begin{equation}\label{eq6}
\hbar\omega_{\max} = \Delta - \frac{8\pi\,e^2|z_{u,l}|^2\Sigma_i(1 + \Delta/\Delta_i)}{\kappa\,d}
\end{equation}
is the inter-GL resonant transition
energy, where the second term in the right-hand side constitutes the depolarization shift(see, for instance, 
Ref.~\cite{29}),
which for media with the population inversion is negative, 
$z_{u,l} = \int\varphi_u^*(z)z\varphi_l(z)$, where $\varphi_u(z)$ and  $\varphi_l(z)$
are the $z-$dependent factors of the wave functions in the upper and lower GLs, 
$\gamma$
is the relaxation broadening, and $\Delta_i = 4\pi\,e^2d\Sigma_i/\kappa$.

Figure~3 shows the $\Delta$ versus $V$ dependences calculated using Eq.~(4) at $\kappa = 4$, $d = 4$~nm, 
and different values of the dopant densities $\Sigma_i$.

At the resonance $\hbar\omega = \hbar\omega_{max}$, Eq. (5) yields
\begin{equation}\label{eq7}
{\rm Re}~\sigma_{zz}(\omega) = - 
\frac{2e^2|z_{u,l}|^2\Sigma_i(1 + \Delta/\Delta_i)}{\hbar}\biggl(\frac{\hbar\omega_{\max}}{\gamma}\biggr).
\end{equation}
Introducing the  gain-overlap factor for the  THz mode in the wave guide under consideration
$\Gamma = [\int_{-L}^{L}dx |E_z(x,0)|^2/\int_{-L}^L\int_{-W}^Wdxdz|E_z(x,0)|^2]$,
where $E_z(x,z)$ is the spatial distribution of the THz electric field in the TM mode
propagating along the MM waveguide (in the $y$ direction) and $2L$ is the length of GLs (distance between t
he side contacts),
the maximum THz gain $g = 4\pi {\rm Re}\sigma\Gamma/c\sqrt{\kappa}$, where $c$ is the speed of light in vacuum, 
can be estimated as

\begin{equation}\label{eq8}
g = 
\frac{8\pi\,e^2|z_{u,l}|^2\Sigma_i(1 + \Delta/\Delta_i)}{\hbar\,c\sqrt{\kappa}}
\biggl(\frac{\hbar\omega_{max}}{\gamma}\biggr)\Gamma.
\end{equation}
%
\begin{figure}[t]\label{Fig.4}
\begin{center}
\includegraphics[width=7.5cm]{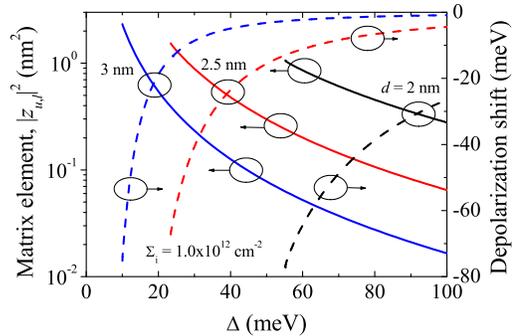}
\caption{Inter-GL matrix element $|z_{u,l}|^2$ (solid lines) and depolarization shift (dashed lines) 
as a function of
gap between the Dirac points in GLs $\Delta$.
}
\end{center} 
\end{figure} 

\begin{figure}[t]\label{Fig.5}
\begin{center}
\includegraphics[width=7.0cm]{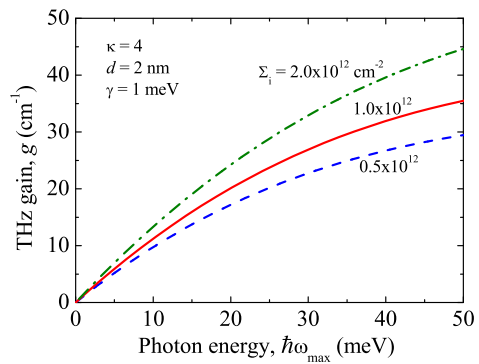}\\
\caption{THz gain $g$ versus the resonant photon energy $\hbar\omega_{max}$ for double-GL
structures with different dopant densities $\Sigma_i$ and $d = 2$~nm.
}
\end{center} 
\end{figure}

\begin{figure}[t]\label{Fig.6}
\begin{center}
\includegraphics[width=7.0cm]{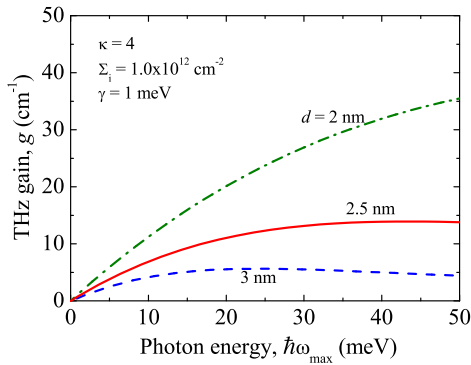}
\caption{The same as in Fig.~5 but for different  $d$ and $\Sigma_i 
= 1.0\times10^{12}$~cm$^{-2}$.
}
\end{center} 
\end{figure}

The value of matrix element $|z_{u,l}|^2$ strongly depends on $d$ and $\Delta$.
It is evaluated using a simple model, in which each GL is considered as 
a delta-layer separated by a   barrier layer made of WS$_2$. Following Ref.~\cite{19}, we assume that  
the conduction band offset between a GL and WS$_2$ barrier and the effective electron mass in WS$_2$ are  
equal to 0.4~eV and 0.27 of the free electron mass, respectively~\cite{29}. 
Figure~4 shows the dependence of the matrix element $|z_{u,l}|^2$ on the energy difference between the 
Dirac points $\Delta$ and the value of the depolarization shift (dependent of
$|z_{u,l}|^2$) for  different thicknesses of the WS$_2$ inter-GL barrier.
As seen from Fig.~4, $|z_{u,l}|^2$ dramatically decreases with increasing $\Delta$. According to 
Eqs.~(5) - (8), this substantially affects the transverse ac conductivity
Re~$\sigma_{zz}(\omega)$, the resonant value of the photon energy $\hbar\omega_{max}$, and, hence 
the THz gain $g$.

Figures~5 and 6 show the resonant THz gain versus photon energy $\hbar\omega_{max}$ 
for different dopant densities $\Sigma_i$ and the inter-GL barrier layer thicknesses.
In Fig.~(5), it is assumed that
 $\kappa = 4$, $d = 2$~nm and $\gamma/\hbar = 1.6\times10^{12}~s^{-1}$. The gain-overlap factor  
 for the MM waveguide with $W = 5~\mu$m is set to be $\Gamma = 10^{3}$~cm$^{-1}$.
As seen from Fig.~5, an increase in the dopant density naturally increases the THz gain. However, 
this increase can
be limited by  an increase 
in broadening parameter $\gamma$.  
One can also see that the THz gain exceeds (or even well exceeds)  the value 10~cm$^{-1}$ in  the range 
$\hbar\omega_{max} \gtrsim 10$~meV ($\omega_{max}/2\pi \gtrsim 2.5$~THz). Hence,
$g$ can exceed the TM-mode losses $\alpha$ in the copper MM waveguides even at room temperature~\cite{31}. 
 It is worth noting that the
Drude absorption for the mode under consideration can be very weak because of relatively
small THz electric field component $E_x$ (as seen from Fig.2,  the ratio $|E_z|/|E_x| > 100$), 
particularly at $z = 0$ , where the double-GL structure is placed. 
The THz gain of the TM mode in the laser under consideration
is at least comparable the maximum THz gain (without the Drude losses) in  the injection lasers utilizing the TE mode 
and the intra-GL radiative transitions~(see, for instance, Ref.~\cite{6}).
Hence the former device can exhibit   advantages due to the effective suppression 
of the Drude absorption.
As follows from Eq.~(8),  the THz gain weakly depends on the temperature. Some temperature dependence 
appears due to the temperature dependence of parameter $\gamma)$ or at low doping levels when $\mu \lesssim T$). 
Weak temperature dependence of the resonant tunneling in double-GL structures was observed 
experimentally~\cite{18} (see also Refs.~\cite{19,20}).
This can provide the superiority of the  double-GL lasers under consideration over the GL-based lasers
using the intra-GL transitions discussed previously~\cite{4,5,6} and QCL lasers~\cite{12,31} at elevated 
temperatures in the low end of the THz range as well as in the range where the
operation of QCLs is hampered by the optical phonon absorption.

The THz gain in the double-GL structures with thicker inter-GL layer is markedly smaller
as seen from Fig.~6.   In this case, to overcome the waveguide losses
the device structure with an active region consisting of  a system of several
parallel double-GL structures can be used. Each the double-GL system can be separated from others by 
relatively thick, non-transparent barrier layer (with the thickness $D \gg d$.
In such a case,the quantity $g$  for the THz gain  given by Eq.~(8) should be multiplied
by the number of the double-GL systems in the device structure.
 The multiple-GL structures in question can provide a much higher net THz gain without a marked increase 
 in the Drude absorption enabling room temperature THz lasing in the range
 $\hbar\omega_{max} \simeq 5 - 10$~meV ($\omega_{max}/2\pi \simeq 1.25  - 2.5$~THz), where
 they might compete with the THz oscillators based on A$_3$B$_5$
 resonant-tunneling structures~\cite{32}.

In conclusions, we proposed THz lasers based on double-GL structures using the inter-GL radiative transitions 
and estimated their THz gain at different photon energies and applied voltages. It was demonstrated that 
the laser can exhibit advantages over other THz lasers
due to the essential suppression of the Drude absorption in GL, weak temperature dependence
of the THz gain even at room temperature, and voltage tuning of the spectrum of the emitted photons.

The authors are grateful to A. Satou and D. Svintsov for useful discussions.
This work was supported by the 
Japan Society for Promotion of Science (Grant-in-Aid for Specially Promoting Research, No.23000008)
and the Japan Science and Technology Agency (CREST Project)), Japan, as well as by  the Russian Foundation 
of Basic Research and the Dynasty Foundation, Russia.

\end{document}